\documentclass[a4paper,twocolumn,aps,eqsecnum,nofootinbib]{revtex4}
\usepackage{graphicx}
\usepackage{bm}
\usepackage{latexsym}
\usepackage{amsmath}
\usepackage{amssymb}
\usepackage{amsfonts}
\usepackage[dvips]{color}
\newcommand{\half}{\frac{1}{2}}

\newcommand{\der}{\partial}
\newcommand{\bra}{\langle}
\newcommand{\ket}{\rangle}
\newcommand{\dsl}{\partial\kern-0.55em\raise 0.14ex\hbox{/}}

\newcommand{\bfx}{\bm{x}}

\newcommand{\intmom}[1]{\int\!\frac{d^{4}#1}{(2\pi)^{4}}}
\newcommand{\intmomd}[1]{\int\!\frac{d^{d}#1}{(2\pi)^{d}}}
\newcommand{\mom}[1]{\frac{d^{4}#1}{(2\pi)^{4}}}
\newcommand{\Tr}{\mathop{\rm Tr}\nolimits}
\newcommand{\tr}{\mathop{\rm tr}\nolimits}
\newcommand{\Det}{\mathop{\rm Det}\nolimits}
\newcommand{\Ln}{\mathop{\rm Ln}\nolimits}

\newcommand{\GITs}{\Gamma_{\mathrm{ITs}}^{(1)}}
\newcommand{\GANTs}{\Gamma_{\mathrm{ANTs}}^{(1)}}
\newcommand{\GK}{\hat{\Gamma}^{(1)}}
\begin{document}

\preprint{KYUSHU-HET-119}

\title{Apparently noninvariant terms of $U(N)\times U(N)$ nonlinear
sigma model in the one-loop approximation}

\author{Koji Harada}
\email{harada@phys.kyushu-u.ac.jp}
\affiliation{Department of Physics, Kyushu University\\
Fukuoka 810-8560 Japan}
\author{Hirofumi Kubo}
\email{kubo@higgs.phys.kyushu-u.ac.jp}
\affiliation{Department of Physics, Kyushu University\\
Fukuoka 810-8560 Japan}
\author{Yuki Yamamoto}
\email{yamamoto@koeki-u.ac.jp}
\affiliation{
Department of Community Service and Science,
Tohoku University of Community Services and Science\\
Iimoriyama 3-5-1, Sakata 998-8580 Japan}

\date{\today}

\begin{abstract}
 We show how the Apparently Noninvariant Terms (ANTs), which emerge in
 perturbation theory of nonlinear sigma models, are consistent with the
 nonlinearly realized symmetry by employing the Ward-Takahashi identity
 (in the form of an inhomogeneous Zinn-Justin equation).  In the
 literature the discussions on ANTs are confined to the $SU(2)$ case. We
 generalize them to the $U(N)$ case and demonstrate explicitly at the
 one-loop level that despite the presence of divergent ANTs in the
 effective action of the ``pions,'' the symmetry is preserved.
\end{abstract}


\maketitle

\section{Introduction}

It has been well known that perturbation theory for nonlinear sigma
models (NLSMs) produces various types of apparently noninvariant terms
(ANTs), when ``pion fields'' are introduced. In a previous
paper~\cite{Harada:2009nb}, we classified them into two kinds: the first
kind of ANTs refers to those that are quartically divergent and do not
vanish in the zero momentum limit. It is well understood that the
contributions from the Jacobian cancel them~\cite{Charap:1971bn,
Honerkamp:1971va, Gerstein:1971fm}. (In the dimensional regularization
scheme, this kind of terms and the contributions from the Jacobian are
both absent.)

The second kind~\cite{Tataru:1975ys, Honerkamp:1971sh, Kazakov:1976tj,
deWit:1979gw, Appelquist:1980ae} is more subtle. They are also
divergent, but do vanish in the zero momentum limit. They appear even
with dimensional regularization, but cannot be absorbed by symmetric
counterterms~\cite{Tataru:1975ys, Appelquist:1980ae}. In
Ref.~\cite{Harada:2009nb}, we employed lattice regularization and
investigated the second-kind ANTs. We have shown that the ANTs emerge
despite the manifest symmetry present in the lattice formulation and
that they have nothing to do with the Jacobian, which is well defined in
this formulation.  The appearance of the second-kind ANTs are thus
consistent with the symmetry of the NLSM. In the following, we
concentrate on the second-kind ANTs and refer to them simply as ANTs.

Natural questions are: \textit{why do ANTs emerge? How are they
consistent with the symmetry?}  In the present paper, we answer these
questions.

In order to consider the symmetry and to investigate its consequences,
it is useful to employ the Ward-Takahashi (WT) identity for the
effective action.  The invariance under the transformation of the fields,
collectively denoted as $\phi(x)$, $\phi(x)\to \phi(x) + \delta
\phi(x)$, can be expressed as
\begin{equation}
 \int d^4x \bra \delta \phi^a(x)\ket 
  \frac{\delta \Gamma[\varphi]}{\delta \varphi^a(x)}
  =0,\label{WT}
\end{equation}
where, $\Gamma[\varphi]$ is the effective action for the classical field
$\varphi^a(x) = \bra \phi^a(x) \ket$.

A crucial observation is that in general $\bra \delta \phi^a(x)\ket$ is
not equal to $\delta \varphi^a(x)$ for a nonlinearly realized
symmetry. Thus at the quantum level, the symmetry transformation for
$\varphi$ is different from the defining transformation law for the
quantum field $\phi$.

From this observation, it is obvious that the appearance of ANTs is not
a phenomenon particular to NLSMs, but is a general feature of the
theories with nonlinearly realized symmetry.

To treat nonlinear transformations, we introduce an external field that
couples to $\delta \phi$, as Zinn-Justin~\cite{ZinnJustin:1974mc} did
for the BRST symmetry of nonabelian gauge theory. The difference between
the present case and the BRST case is that the BRST symmetry is
nilpotent while the present symmetry is not. Thus, the resultant WT identity
has an inhomogeneous term in the present case.

Furthermore, in order to express the WT identity in a compact form, we
need to make a special choice for the ``pion'' fields. In the case of
$SU(2) \times SU(2)$ NLSM, a useful parameterization of the field
$U(x) \in SU(2)$ is
\begin{equation}
 U(x) = \frac{1}{F}
  \left[
   \sigma(x) +i\vec{\tau}\cdot \vec{\pi}(x)
  \right], 
  \quad 
  \sigma(x)\equiv \sqrt{F^2-\vec{\pi}^2(x)}.
  \label{su2sqrt}
\end{equation}
Under the infinitesimal $SU(2)\times SU(2)$ transformation, these fields
transform as follows,
\begin{eqnarray}
 \delta \sigma &=& -\vec{\beta} \cdot \vec{\pi},
  \\
 \delta \vec{\pi} &=& -\vec{\alpha}\times \vec{\pi} + \vec{\beta}\sigma,
\end{eqnarray}
where $\vec{\alpha}$ and $\vec{\beta}$ are parameters for the vector and
the axial-vector transformations. The point is that a finite number of
fields ($\vec{\pi}$ and $\sigma$ in the present case) form a closed set
under the transformation (i.e., a multiplet), in the sense that they
transform mutually among them. This property is absent for a generic
parameterization. For example, in the case of the exponential
parameterization, $U(x) = \exp ( i \vec{\tau} \cdot \vec{\pi} (x) / F)$,
the axial-vector transformation law for the so-defined pion field
introduces another operator $\Sigma^{ab}\equiv\tr \left[\tau^a U^{-1}
\tau^b U +\tau^a\tau^b\right]$,
\begin{equation}
 \delta \pi^a =
  \frac{F}{4}\beta^b\Sigma^{ab}.
\end{equation}
If one transforms $\Sigma^{ab}$, the result cannot be expressed as a
linear combination of $\vec{\pi}$ and $\Sigma^{ab}$, and introduces
another operator. No finite sequence of this procedure does form a
closed set under the axial-vector transformation.

A special parameterization such as Eq.~(\ref{su2sqrt}) for the $SU(2)
\times SU(2)$ case is desired also for a general group, but it is not
obvious to find one for a group other than $SO(N)$. (Note that $SU(2)
\times SU(2) \approx SO(4)$.)

In this paper, we consider $U(N)\times U(N)$ NLSMs and formulate the WT
identity in the form of an inhomogeneous Zinn-Justin
equation\footnote{For a small value of $N$, e.g., $N=2$ or $3$, there
are nontrivial relations which reduce the number of independent terms
to be discussed in this paper. It however does not affect the main
conclusions.}. By using it, we investigate how the WT identity is
satisfied despite the presence of ANTs in the one-loop approximation. We
determine the form of the divergent part of the ANTs to all order in
$\varphi$. We also show that ANTs of the six-point proper vertices (the
1PI parts of the amputated connected Green function) do not vanish on
shell.

Several comments on the literature are in order. 

(i) T\u{a}taru~\cite{Tataru:1975ys} calculates the effective action for
the $SO(N)$ NLSM and shows that the ANTs are proportional to $\delta
S/\delta \varphi$ at the one-loop level, where $S$ is a classical
action. It implies that there is a field redefinition that eliminates
ANTs. It thus implies that the ANTs do not contribute to the S matrix
elements in the one-loop approximation.

(ii) Appelquist and Bernard~\cite{Appelquist:1980ae} emphasize that the
ANTs can be eliminated by a field redefinition and thus the symmetry is
not lost. They however do not examine the WT identity. Their statement
may be translated as follows in terms of the effective action: let us
assume that the field $\varphi(x)$ can be expressed as a function of a
new field $\varphi'$ and its derivatives\footnote{The definition of the
new field must involves the derivatives because T\u{a}taru actually
shows that the existence of ANTs is independent of the choice of
parameterization. A redefinition of the field without derivatives just
gives another parameterization.}, $\varphi =\varphi(\varphi', \der
\varphi', \cdots)$. Their statement is that, with a suitable choice of
$\varphi'$, the effective action in term of $\varphi'$,
\begin{equation}
 \Gamma'[\varphi'] = 
  \left.
   \Gamma[\varphi]
  \right|_{\varphi=\varphi(\varphi', \der \varphi',\cdots)},
\end{equation}
is invariant under the \textit{naive} transformation law\footnote{The
naive transformation law for the classical field means that it
transforms in the same way as the corresponding quantum field.} for
${\varphi'}^a$,
\begin{equation}
 \int d^4x\; \delta {\varphi'}^a(x)
  \frac{\delta \Gamma'[\varphi']}{{\varphi'}^a(x)}=0.
  \label{trfdWT}
\end{equation}
Note that, unlike the field $\varphi$, the new field $\varphi'$ does not
have a direct connection to the quantum field $\phi$, and thus the
relation between (\ref{WT}) and (\ref{trfdWT}) is obscure. The
definition of the new field in terms of the old one may be found only
after calculating the ANTs.

(iii) Brezin, Zinn-Justin, and Le Guillou~\cite{Brezin:1976ap} consider
the $SU(2) \times SU(2)$ NLSM in $2+\epsilon$ dimensions. They examine
the symmetry with the help of the inhomogeneous Zinn-Justin equation. In
this respect, their work is very close to our present work. As they
noted the use of the special parameterization (\ref{su2sqrt}) is
essential for the formulation of the Zinn-Justin equation. In two
dimensions, the NLSM is renormalizable if the Lagrangian does not
contain the higher derivatives of the fields than two. In such a case,
the field redefinition that eliminates ANTs does not depend on the
derivatives and is nothing but a usual wavefunction renormalization.
They can also obtain the most general form of the ANTs in this case.
But it is a very special situation. If one includes more derivatives in
the Lagrangian, the theory is not renormalizable, and the field
redefinition inevitably involves derivatives. See also
Ref.~\cite{Bardeen:1976zh}.

(iv) Recently Ferrari and his collaborators~\cite{Ferrari:2005ii,
Ferrari:2005va, Ferrari:2005fc, Bettinelli:2007zn} discuss the $SU(2)
\times SU(2)$ NLSM in perturbation theory based on a local functional
equation, which is nothing but the Schwinger-Dyson equation in the
functional form.  Although our work looks similar to theirs in the respect
that both utilize functional identities, there are important differences
between them. First of all, they try to construct a theory with only two
constants,
while we
regard the NLSM as an effective field theory, that is, a
nonrenormalizable theory with infinitely many higher dimensional
operators. Secondly, their functional identity is a local one, while our
analysis is based on the usual WT identity for rigid
transformations. Thirdly (and perhaps most importantly), they subtract
ANTs as well as symmetric divergences, while we only subtract
divergences occurring in apparently invariant terms keeping the ANTs
intact. 

We consider the loop-wise expansion in order to examine how the WT
identity is satisfied. At the tree level, the effective action is
nothing but a classical action, and the invariance is trivial,
irrespective to the number of derivatives it contains. ANTs emerge only
in loop corrections.

For an effective theory with infinitely many terms with an increasing
number of derivatives, it is impossible to calculate all the one-loop
corrections. Fortunately, however, because of the chiral symmetry, the
NLSMs admit a derivative expansion~\cite{Weinberg:1978kz,
Gasser:1983yg}. To $\mathcal{O}(p^4)$, to which order the one-loop
corrections start to contribute, there are only a finite number of
independent operators.

Although the main result of the present paper is the clarification of
the consistency of ANTs with the underlying symmetry by the extensive
use of the inhomogeneous Zinn-Justin equation in the one-loop
approximation, we also clarify several confusing points in the
literature; (i) The counterterms for ANTs break the symmetry therefore
should not be added. The resulting effective action contains apparently
noninvariant divergences which however do not contribute to the
S-matrix. (ii) No field redefinition is needed to make the theory
invariant. (iii) ANTs for the proper vertices in general do not vanish
even on shell.

The investigation of ANTs would have crucial importance in the Wilsonian
renormalization group (RG) analysis of theories with nonlinearly
realized symmetries. RG transformations generate all the terms
consistent with the symmetry, thus ANTs as well. One needs to understand
ANTs well to obtain correct RG equations. An interesting example is a
Wilsonian RG analysis of the chiral perturbation theory with nucleons,
which would require proper treatment of ANTs. We expect that it gives
dynamical explanations for nonperturbative aspects of the theory such as
resonances.

The structure of the paper is the following: In Sec.~\ref{U_N}, we
introduce the special parameterization of the $U(N) \times U(N)$ NLSM
according to Ref.~\cite{Gasiorowicz:1969kn}, which allows us to
formulate the WT identity in the form of an inhomogeneous Zinn-Justin
equation. In Sec.~\ref{one-loop}, an explicit one-loop calculation of
$\mathcal{O}(\varphi^3)$ is shown, which illustrates how the WT identity
is satisfied despite the presence of ANTs up to including
$\mathcal{O}(p^4)$.  The exact form of the divergent part of the ANTs of
the one-loop effective action is determined to all order in $\varphi$ by
using the invariance argument.  In Sec.~\ref{six-point}, we show that
the ANTs of the six-point proper vertices do not vanish on shell, while
the six-point amputated connected Green functions do vanish on
shell. Finally in Sec.~\ref{summary}, we summarize the
results. Appendix~\ref{group} collects some formulae for $U(N)$ which
are useful in simplifying the results.  We present the results for the
$SU(2) \times SU(2)$ case in Appendix~\ref{sec:su2} to compare them with
those in the literature.

\section{WT identity for $U(N)\times U(N)$ NLSMs}
\label{U_N}

As we discussed in the previous section, in order to formulate the WT
identity in the form of a Zinn-Justin equation, it is necessary to have
a special parameterization of the field. Interestingly, it is possible by
introducing the ``parity doublet'' for parameterizing
$U(N)$~\cite{Gasiorowicz:1969kn}.

Consider $U(x)$ as
\begin{equation}
 U(x)=\frac{1}{F}\left(\sigma(x) + i\phi(x)\right),
  \label{Udef}
\end{equation}
where $\sigma(x)$ and $\phi(x)$ are Hermitian $N\times N$ matrices, and
can be expressed by using the generators of $U(N)$, $T_i, \ (i=0,1,
\cdots, N^2-1$):
\begin{equation}
 \sigma(x)=2\sigma_{i} T_{i}, \quad \phi(x) =2\phi_{i} T_{i}.
\end{equation}
The generators $T_i$ are normalized as
\begin{equation}
 \tr\left(T_{i}T_{j}\right)=\half \delta_{ij},
\end{equation}
and satisfy the following relations,
\begin{equation}
 [T_{i},T_{j}]=if_{ijk}T_{k}, \quad  \{T_{i},T_{j}\}=d_{ijk}T_{k},
  \label{U_N_algebra}
\end{equation}
where $f_{ijk}$ is a completely antisymmetric structure constant and
$d_{ijk}$ is a completely symmetric tensor.  Note in particular that
$T_0$ is proportional to the identity matrix,
\begin{equation}
  T_{0}=\frac{1}{\sqrt{2N}}I, \label{T0}
\end{equation}
and 
\begin{equation}
  f_{0jk}=0, \quad  d_{0jk}=\sqrt{\frac{2}{N}}\delta_{jk}.
   \label{f_and_d}
\end{equation}

The $U(N)\times U(N)$ transformation law for $U$ is given by
\begin{equation}
 U \to g_L U g_R^\dagger, \label{UN_UN}
\end{equation}
where $g_L$ and $g_R$ are elements of $U(N)_L$ and $U(N)_R$ respectively.
Under the infinitesimal transformation, $\sigma$ and $\phi$ transform as
\begin{eqnarray}
  \delta\sigma_{i}&=&-f_{ijk}\alpha_{j}\sigma_{k}
   -d_{ijk}\beta_{j}\phi_{k},
   \\
  \delta\phi_{i}&=&-f_{ijk}\alpha_{j}\phi_{k}
   +d_{ijk}\beta_{j}\sigma_{k},
\end{eqnarray}
where $\alpha_i$ and $\beta_i$ are the parameters for the $U(N)_V$ and
$U(N)_A$ transformations respectively.  Note that successive
transformations of $\sigma_i$ and $\phi_i$ can be written as linear
combinations of $\sigma_k$ and $\phi_k$.

Under the parity transformation, $U(t,\bfx)$ transforms to
$U^\dagger(t,-\bfx)$, so that $\sigma(t,\bfx)$ to $\sigma(t,
-\bfx)$ and $\phi(t, \bfx)$ to $-\phi(t, -\bfx)$.

For the $N=2$ case, $(\sigma_0, \vec{\phi})$ corresponds to $\sigma$ and
$\vec{\pi}$ in Eq.~(\ref{su2sqrt}), while $(\phi_0, -\vec{\sigma})$ is
its parity partner.

In order to obtain the NLSM, one needs to impose the constraints. To do
so consistently, we consider the path-integral measure. Before imposing
the constraints, the measure should be 
\begin{equation}
 \mathcal{D}U\mathcal{D}U^\dagger = \mathcal{D}\sigma\mathcal{D}\phi.
\end{equation}
Here and hereafter we ignore the irrelevant normalization factor. The
constraints we impose are
\begin{equation}
 U^\dagger U=\frac{1}{F^2}
  \left(\sigma^2 +i\left[\sigma,\phi\right]+\phi^2\right)=I, 
  \quad \left[\sigma,\phi\right]=0.
\end{equation}
Note that these constrains are invariant under $U(N)\times U(N)$
transformation Eq.~(\ref{UN_UN}). They are to be inserted as delta
functions:
\begin{equation}
 \mathcal{D}\sigma\mathcal{D}\phi\delta (\sigma^2+\phi^2-F^2 I)
  \delta\left(\left[\sigma,\phi\right]\right).
\end{equation}
The first delta function may be written as
\begin{eqnarray}
 \delta(\sigma^2+\phi^2-F^2 I)
  &=&
  \frac{1}{\det\left[2d_{ijk}\sigma_k(\phi)\right]}
  \nonumber \\
 &&\times
  \bigg[
  \delta\left(\sigma -\sigma(\phi)\right)
  +\delta\left(\sigma+\sigma(\phi)\right)
  \bigg],\nonumber \\
\end{eqnarray}
where $\sigma\left(\phi\right)=\sqrt{F^2-\phi^2}$ is a matrix, and is
defined as a power series of $\phi^2$,
\begin{equation}
 \sigma(\phi)\equiv
  \sqrt{F^2-\phi^2}=F-\frac{1}{2F}\phi^2-\frac{1}{8F^3}\phi^4 + \cdots.
  \label{sigma}
\end{equation}
and since we are going to work in perturbation theory (i.e., in the
vicinity of $U(x)=I$), we only consider
the first delta function. 

The second constraint $\left[\sigma, \phi\right]=0$ is now automatically
satisfied. Thus the measure with the constraints reduces to
\begin{equation}
 \frac{\mathcal{D}\phi}
{\det\left[d_{ijk}\sigma_k(\phi)\right]},
\end{equation}
where we have dropped $\delta (0)$ from the second constraint as an
irrelevant constant. Since we are going to work with the dimensional
regularization, the determinant factor does not give nontrivial
contributions, and we ignore it hereafter. Our experience with the
lattice calculation~\cite{Harada:2009nb} leads us to believe that this
factor has nothing to do with ANTs. 

Note that the field $U$ introduced in Eq.~(\ref{Udef}) originally has
$2N^2$ real components but now it has $N^2$ real components: $N^2-1$ for
the Nambu-Goldstone (NG) bosons associated with the broken $SU(N)_A$
symmetry and one for the NG boson associated with the broken
$U(1)_A$. If one wishes to incorporate the axial anomaly and the masses,
one needs to introduce explicit symmetry breaking terms, as we do in the
effective theory of QCD.  In this paper, however, we do not consider
such breaking terms for simplicity.

The generating functional in Euclidean space is now given by
\begin{equation}
 e^{-W[J, K]}\!=\!\int\! \mathcal{D}\phi \;
  \exp
  \left[
   -\!\int\! d^4x \left(\mathcal{L} +\sigma_iK_i \right)
   \!+\!\int\! d^4x \phi_iJ_i
  \right],
\end{equation}
where $\sigma_i(x)$ is actually given as in Eq.~(\ref{sigma}). The
Lagrangian is now for a general $U(N) \times U(N)$ invariant NLSM,
\begin{equation}
 \mathcal{L}=\frac{F^2}{4}\tr\left(\der_\mu U^\dagger \der_\mu U\right)
  + \cdots,
  \label{Lagrangian_to_p^2}
\end{equation}
where the ellipsis denotes the terms with more than two derivatives, and
$U(x)$ is written in terms of $\phi(x)$ as in Eq.~(\ref{Udef}) with
Eq.~(\ref{sigma}). The explicit forms of the terms of $\mathcal{O}(p^4)$
will be shown in Eqs.~(\ref{L1}) -- (\ref{L6}) in Sec.~\ref{one-loop}.

As usual, we introduce the effective action, 
\begin{equation}
 \Gamma[\varphi, K]=W[J, K]+\int d^4x J_i\varphi_i\; ,
\end{equation}
where $\varphi(x)$ is the expectation value of $\phi(x)$ in the presence
of the external fields $J(x)$ and $K(x)$,
\begin{equation}
 \varphi_i(x)=-\frac{\delta W[J,K]}{\delta J_i(x)}
  =\bra \phi_i(x)\ket_{J,K} \; .
\end{equation}

Invariance of 
\begin{equation}
 \mathcal{D}\phi \; \exp\left[-\int d^4x\; \mathcal{L}\right]
\end{equation}
under the infinitesimal axial-vector transformation leads to the
following WT identity,
\begin{equation}
 \int d^4x 
  \left[
   \bra \delta_A \sigma_i(x)\ket_{J,K} K_i(x)
   -\bra \delta_A \phi_i(x)\ket_{J,K} J_i(x)
  \right]=0.
\end{equation}
Inserting 
\begin{eqnarray}
 \bra \delta_A \sigma_i\ket_{J,K}\!\!&=&\!
  -d_{ijk}\beta_j\bra \phi_k\ket_{J,K}
  =-d_{ijk}\beta_j\varphi_i,\\
 \bra \delta_A \phi_i\ket_{J,K}\!\!&=&\!
  d_{ijk}\beta_j\bra \sigma_k\ket_{J,K}
  =d_{ijk}\beta_j\frac{\delta \Gamma[\varphi, K]}{\delta K_i},\
\end{eqnarray}
and
\begin{equation}
 J_i=\frac{\delta \Gamma[\varphi, K]}{\delta \varphi_i}
\end{equation}
into the WT identity, we arrived at the following inhomogeneous
Zinn-Justin equation,
\begin{equation}
 d_{ijk}\int d^4x 
  \left(
   \frac{\delta \Gamma[\varphi, K]}{\delta K_j(x)}
   \frac{\delta \Gamma[\varphi, K]}{\delta \varphi_k(x)}
   +\varphi_j(x) K_k(x)
  \right)=0.
  \label{ZinnJustin}
\end{equation}
Note that this is not a local equation. If we consider a local
axial-vector transformation function $\beta_j(x)$, we would arrive at a
local equation which however contains the expectation value of the
divergence of the axial-vector current as an additional term. The
additional term cannot be written in terms of the derivatives of the
effective action with respect to $\varphi$ and $K$. Furthermore, since
we are considering an effective theory with infinitely many terms, the
axial-vector current is not just $\sim \tr(U^\dagger\der_\mu U-U\der_\mu
U^\dagger)$ but depends on infinitely many terms with derivatives. We
find that a local version is not useful.

The effective action $\Gamma[\varphi, K]$ may have a loop-wise expansion:
\begin{equation}
   \Gamma[\varphi,K]=\sum_{n=0}\Gamma^{(n)}[\varphi,K],
\end{equation}
where the zeroth order term is the classical action,
\begin{equation}
 \Gamma^{(0)}[\varphi,K]=S[\varphi,K]\equiv 
  \int d^4x \left(\mathcal{L}(\varphi)+\sigma_i(\varphi) K_i\right),
  \label{classical_action}
\end{equation}
where $\sigma(\varphi)$ is a function of the classical field, $\varphi$,
with the definition Eq.~(\ref{sigma}).

At the zeroth order (at the tree level), Eq.~(\ref{ZinnJustin}) gives
\begin{equation}
  d_{ijk}\int d^4x 
  \left(
   \frac{\delta S[\varphi, K]}{\delta K_j(x)}
   \frac{\delta S[\varphi, K]}{\delta \varphi_k(x)}
   +\varphi_j(x) K_k(x)
  \right)=0.
\end{equation}
With the definition Eq,~(\ref{classical_action}), it is
\begin{equation}
 d_{ijk}\int d^4x 
   \sigma_j(\varphi)
   \frac{\delta S[\varphi]}{\delta \varphi_k(x)}
   =0\; ,
   \label{cls_inv}
\end{equation}
which just implies the invariance of the classical action,
$S[\varphi]\equiv \int d^4x \mathcal{L}(\varphi)$ under the axial-vector
transformation, $\delta_A \varphi_i=d_{ijk}\beta_j
\sigma_k(\varphi)$. Note that in deriving Eq.(\ref{cls_inv}), we have
used
\begin{equation}
 \delta_A \sigma_i(\varphi) 
  =\frac{\der \sigma_i}{\der \varphi_j}\delta_A \varphi_j.
\end{equation}

At the first order (at the one-loop level), Eq.~(\ref{ZinnJustin}) gives
\begin{align}
 &d_{ijk}
  \int d^{4}x\;
  \bigg(
  \frac{\delta S[\varphi,K]}{\delta K_{j}(x)}
  \frac{\delta \Gamma^{(1)}[\varphi,K]}{\delta \varphi_{k}(x)}
  \nonumber \\
 &\qquad \qquad \qquad {}+
  \frac{\delta \Gamma^{(1)}[\varphi,K]}{\delta K_{j}(x)}
  \frac{\delta S[\varphi,K]}{\delta \varphi_{k}(x)}
  \bigg)
  =0.
 \label{one-loop-ZJ}
\end{align}
This can be expanded in powers of the external field $K$. At the zeroth
order of this expansion, we have
\begin{align}
 &d_{ijk}
    \int d^{4}x
    \bigg(
 \sigma_j(\varphi)(x)
 \left.\frac{\delta \Gamma^{(1)}[\varphi,K]}{\delta \varphi_{k}(x)}\right|_{K=0}
 \nonumber \\
 &\qquad \qquad \qquad {}+
 \left.\frac{\delta \Gamma^{(1)}[\varphi,K]}{\delta K_{j}(x)}\right|_{K=0}
 \frac{\delta S[\varphi]}{\delta \varphi_{k}(x)}
 \bigg)
 =0.
 \label{oneloopZJ}
\end{align}
This is an important equation. The first term is the variation of the
one-loop contribution to the effective action $\Gamma^{(1)}[\varphi]
\equiv \left.\Gamma^{(1)}[\varphi, K]\right|_{K=0}$ under the naive
axial-vector transformation. This equation tells us that the first term
does not need to vanish. The symmetry requires that the \textit{sum} of
these two terms should vanish, but not necessarily individually. If the
first term does not vanish, we see there is an ANT.

Another important point is that the apparent noninvariance of the
one-loop contribution to the effective action $\Gamma^{(1)}[\varphi]$ is
proportional to $\delta S/\delta \varphi$. With the use of the equations
of motion, $\Gamma^{(1)}[\varphi]$ is invariant under the naive
transformation.  Note that we have reached this conclusion without doing
any explicit calculations.

In the next section, we demonstrate the explicit calculation of
$\Gamma^{(1)}[\varphi,K]$, and show how the Eq.~(\ref{oneloopZJ}) is
satisfied for the first few terms in the expansion in powers of $\varphi$.

\section{An explicit one-loop calculation}
\label{one-loop}

In this section, we present a detailed one-loop calculation for the
effective action of the $U(N) \times U(N)$ NLSM. As we explained in
Introduction, we also do a low-momentum expansion and calculate the
effective action up to $\mathcal{O}(p^4)$. Only the vertices of
$\mathcal{O}(p^2)$ (obtained by expanding the terms explicitly shown in
Eq.~(\ref{Lagrangian_to_p^2}) in powers of $\phi$) can enter one-loop
diagrams at this order\footnote{This refers to the case with dimensional
regularization. In the cutoff scheme such as lattice regularization,
the loop-wise expansion does not match the expansion in powers of momenta.
}.

\subsection{Preliminary: expansion in powers of $\varphi$}

We are going to show that the effective action has a part which are not
apparently symmetric under the axial-vector transformation. It is useful
to divide $\Gamma^{(1)}[\varphi, K]$ into three parts,
\begin{equation}
 \Gamma^{(1)}[\varphi,K]=\GITs[\varphi]+\GANTs[\varphi]+\GK[\varphi,K]
  \; ,
  \label{eqn:Gamma-divide}
\end{equation}
where $\GITs[\varphi]$ is the apparently invariant part satisfying
\begin{equation}
 d_{ijk}\int d^4x 
   \sigma_j(\varphi)
   \frac{\delta \GITs[\varphi]}{\delta \varphi_k(x)}
   =0\; ,
\end{equation}
and $\GANTs[\varphi]$ is the apparently noninvariant part. All the
$K$-dependent terms are contained in $\GK[\varphi,K]$, which vanishes
when we set $K=0$.

There is a potential ambiguity in the separation into the apparently
``invariant'' and ``noninvariant'' parts. Even if one subtracts an
invariant piece from the ``invariant'' part and adds it to the
``noninvariant'' one, they are still ``invariant'' and ``noninvariant''
parts respectively. This ambiguity does not affect the following
discussion in any significant way.

Let us define the momentum space quantities in $d$ dimensions as follows:
\begin{eqnarray}
 \sigma_{j}(p)&\equiv&
  (2\pi)^{d}\left.\frac{\delta S[\varphi,K]}{\delta K_{j}(-p)}\right|_{K=0}
  \\
 A_{k}(p)&\equiv&
  (2\pi)^{d}\frac{\delta \GITs[\varphi]}{\delta \varphi_{k}(p)}
  \\
 B_{k}(p)&\equiv&
  (2\pi)^{d}\frac{\delta \GANTs[\varphi]}{\delta \varphi_{k}(p)}
  \\
 C_{j}(p)&\equiv&
  (2\pi)^{d}
  \left.\frac{\delta \GK[\varphi,K]}{\delta K_{j}(p)}\right|_{K=0}
  \\
 D_{k}(p)&\equiv&
  (2\pi)^{d}\frac{\delta S[\varphi]}{\delta \varphi_{k}(-p)},
\end{eqnarray}
where $\varphi_i(p)$ and $K_i(p)$ are defined as 
\begin{equation}
  \varphi_{i}(x)=\intmomd{p}\varphi_{i}(p)e^{ip\cdot x},
  \ 
  K_{i}(x)=\intmomd{p} K_{i}(p)e^{ip\cdot x}.
\end{equation}

In terms of them, Eq.~(\ref{oneloopZJ}) is now written as
\begin{equation}
  d_{ijk}\intmomd{p}
   \left\{
    \sigma_{j}(p)B_{k}(p)
    +C_{j}(p)D_{k}(p)
   \right\}=0\; .
  \label{eqn:Zinn-Justin-1loop-K0-mon2}
\end{equation}
Let us expand the each quantity in powers of $\varphi$ as follows:
\begin{eqnarray}
  \sigma_{j}(p)&=&\sigma_{j}^{(\varphi^{0})}(p)
   +\sigma_{j}^{(\varphi^{2})}(p)+\cdots,\\
  B_{k}(p)&=&B_{k}^{(\varphi^{3})}(p)+B_{k}^{(\varphi^{5})}(p)+\cdots,\\
  C_{j}(p)&=&C_{j}^{(\varphi^{2})}(p)+C_{j}^{(\varphi^{4})}(p)+\cdots,\\
  D_{k}(p)&=&D_{k}^{(\varphi^{1})}(p)+D_{k}^{(\varphi^{3})}(p)+\cdots.
\end{eqnarray}
Note that $B_{k}^{(\varphi^{1})}(p)=0$, because, with the dimensional
regularization, the one-loop contributions to the two-point function
vanish when $K=0$.

\subsection{Cubic terms in $\varphi$}

The first nontrivial order in the expansion of the left-hand side of
Eq.~(\ref{eqn:Zinn-Justin-1loop-K0-mon2}) is cubic in $\varphi$,
\begin{equation}
  d_{ijk}\intmomd{p}
   \left\{
    \sigma_{j}^{(\varphi^{0})}(p)
    B_{k}^{(\varphi^{3})}(p)
    +
    C_{j}^{(\varphi^{2})}(p)
    D_{k}^{(\varphi^{1})}(p)
   \right\}.
   \label{cubic_in_varphi}
\end{equation}
We immediately get
\begin{eqnarray}
 \sigma_{j}^{(\varphi^{0})}(p)&=&
  \sqrt{\frac{N}{2}} 
  \mu^{-\varepsilon}F\delta_{0j}(2\pi)^{d}\delta^{d}(p)\; , \label{sigma0}\\
 D_{k}^{(\varphi^{1})}(p)&=&
   p^{2}\varphi_{k}(p)\; ,
\end{eqnarray}
where $\mu$ is an arbitrary mass scale\footnote{In going to $d$
dimensions, the coupling constant $F$ should be replaced by
$\mu^{-\varepsilon} F$ everywhere so that the mass dimension of $F$ is
fixed to be one.}, and $\varepsilon\equiv 2-{d}/{2}$. What we need are
the terms in $\Gamma^{(1)}[\varphi, K]$ that are $\mathcal{O}(\varphi^4
K^0)$ and $\mathcal{O}(\varphi^2 K^1)$.

The one-loop contributions to the effective action
$\Gamma^{(1)}[\varphi, K]$ are given as the sum of the 1PI diagrams in
the presence of the external field $\varphi$ and $K$. It is given by 
\begin{eqnarray}
 \Gamma^{(1)}[\varphi, K] \!&=&\!\! \half \ln \Det[D_F^{-1}]
  =
  \frac{1}{2}\Tr\Ln[D_{F}^{-1}]
  \nonumber\\
  &=&\!{}
  -\frac{1}{2}\!\intmomd{k}\frac{1}{k^{2}}G_{ii}(k,-k)
  \nonumber \\
 &&\!\!{}
  -\frac{1}{4}\!\intmomd{k}
  \frac{1}{k_{1}^{2}}G_{ij}(k_{1},-k_{2})
  \frac{1}{k_{2}^{2}}G_{ji}(k_{2},-k_{1})
  \nonumber \\
 &&\!\!{}+ \mathcal{O}(G_{ij}^3)\; ,
  \label{logdet}
\end{eqnarray}
where $D_F$ and $G_{ij}$ are defined as
\begin{eqnarray}
  &&\int d^{d}x\int d^{d}y
  [D_{F}^{-1}(x,y)]_{ij}
  e^{-iq_{1}\cdot x-iq_{2}\cdot y} 
  \nonumber \\
  &=&
  (2\pi)^{d}\frac{\delta^{2}S[\varphi,K]}
  {\delta\varphi_{i}(-q_{1})\delta\varphi_{j}(-q_{2})}
  \nonumber\\
  &=&
  q_{1}^{2}(2\pi)^{d}\delta^{d}(q_{1}+q_{2})\delta_{ij}
  -G_{ij}(q_{1},q_{2}).
\end{eqnarray}
Note that $G_{ij}$ is the field dependent part of the inverse propagator.

The classical action $S[\varphi, K]$ (\ref{classical_action}) consists
of two parts; the part independent of $K$ and the part linear in $K$. We
therefore divide $G_{ij}$ accordingly,
\begin{equation}
  G_{ij}(q_{1},q_{2})
  \equiv
  G_{0,ij}(q_{1},q_{2})
  +
  G_{1,ij}(q_{1},q_{2}).
\end{equation}

Note that the first term of Eq.~(\ref{logdet})
vanishes because of the property of the dimensional regularization. 

\begin{figure}[t]
 \includegraphics[width=0.7\linewidth,clip]{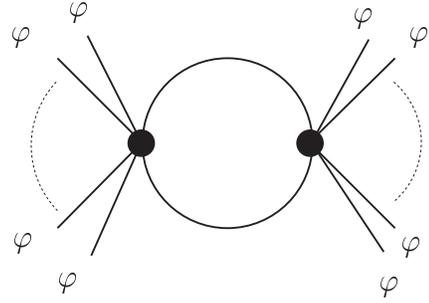} 
\caption{\label{phiphi}The graphical representation of the term
 explicitly shown in Eq.(\ref{zeroK}), the contribution to the effective
 action to the zeroth order in $K$, starting with $\varphi^4$. The blob
 stands for the vertex $G_{0,ij}$, which contains an even number of
 $\varphi$'s.}
\end{figure}

At the zeroth order in $K$, we have
\begin{eqnarray}
 && \frac{1}{2}\ln\Det[D_{F}^{-1}]_{K^{0}} 
  \nonumber \\
  &=&\!\!
  -\frac{1}{4}\! 
  \int 
  \left[\prod_{i=1}^2\frac{d^dk_i}{(2\pi)^d}\right]
  \frac{1}{k_{1}^{2}}G_{0,ij}(k_{1},-k_{2})
  \frac{1}{k_{2}^{2}}G_{0,ji}(k_{2},-k_{1})
  \nonumber \\
 &&+ \mathcal{O}(\varphi^6)\; ,
  \nonumber\\
 \label{zeroK}
\end{eqnarray}
where $G_{0,ij}$ starts with the terms quadratic in $\varphi^2$. See
FIG.~\ref{phiphi}.
\begin{figure}[t]
 \includegraphics[width=0.7\linewidth,clip]{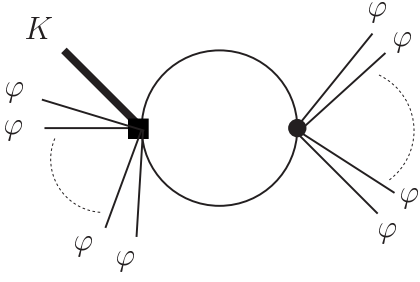}
\caption{\label{Kphi}The graphical representation of the second term
explicitly shown in Eq.(\ref{first_order_in_K}), the contribution to the
effective action to the first order in $K$, starting with
$\varphi^2$. The blob stands for the vertex $G_{0,ij}$, which contains
an even number of $\varphi$, and the square for the vertex $G_{1,ji}$,
which contains a single $K$ as well as zero or an even number of $\varphi$'s.}
\end{figure}

A straightforward calculation gives the following expression for the
first term,
\begin{eqnarray}
 &&-\frac{\bar{\varepsilon}^{-1}}{16F^{4}(4\pi)^{2}}
  \!\int \!\left[\prod_{i=1}^4\frac{d^{4} p_i}{(2\pi)^{4}}\right]
  \!(2\pi)^{4}\delta^{4}\!\left(\sum_{i=1}^4p_{i}\!\right)
  \!\prod_{i=1}^4 \varphi_{l_{i}}(p_{i})
  \nonumber\\
  &&
  \times
  \bigg\{
    d_{kij}d_{kl_{1}l_{2}}d_{gji}d_{gl_{3}l_{4}}
    (p_{1}^{2}+p_{2}^{2})
    (p_{3}^{2}+p_{4}^{2})
    \frac{\mathstrut}{}
  \nonumber\\
  &&\hspace{1em}
  {}+
  2d_{kij}d_{kl_{1}l_{2}}
  (d_{gji}d_{gl_{3}l_{4}}\!\!-\!d_{gjl_{3}}d_{gil_{4}})
  (p_{1}^{2}+p_{2}^{2})
  (p_{3}\cdot p_{4})
  \nonumber \\
 &&\hspace{1em}
  {}+
  2d_{gji}d_{gl_{3}l_{4}}
  (d_{kij}d_{kl_{1}l_{2}}\!\!-\!d_{kil_{1}}d_{kjl_{2}})
  (p_{3}^{2}+p_{4}^{2})
  (p_{1}\cdot p_{2})
  \nonumber\\
  &&\hspace{1em}
  {}+
  4(d_{kij}d_{kl_{1}l_{2}}-d_{kil_{1}}d_{kjl_{2}})
  (d_{gji}d_{gl_{3}l_{4}}-d_{gjl_{3}}d_{gil_{4}})
  \nonumber \\
 &&\hspace{2em}
  {}\times
  (p_{1}\cdot p_{2})(p_{3}\cdot p_{4})
  \nonumber\\
  &&\hspace{1em}
   -\frac{4}{3}
    d_{kil_{1}}d_{kjl_{2}}d_{gjl_{3}}d_{gil_{4}}  
    \nonumber \\
 &&\hspace{2em}
  {}\times
    \left[
      (p_{1}\cdot p_{4})(p_{2}\cdot p_{3})
      -
      (p_{1}\cdot p_{3})(p_{2}\cdot p_{4})
    \right]
  \bigg\}
  \nonumber\\
  &&
  +\text{(finite terms)}\; ,
  \label{eqn:G0-2}
\end{eqnarray}
where $\bar{\varepsilon}^{-1}\equiv  \varepsilon^{-1}-\gamma+\ln 4\pi$.

The next task is to separate the contributions which can be cancelled by
symmetric counterterms from those which cannot be. The symmetric terms
are already contained in the Lagrangian $\mathcal{L}[\phi]$ in
Eq.~(\ref{Lagrangian_to_p^2}), so that the addition of the counterterms
is absorbed by the change of the coefficients of these operators.  We
list all of those terms up to including $\mathcal{O}(p^4)$ that are
invariant under axial-vector transformations. There are six independent
ones;
\begin{eqnarray}
 \mathcal{L}_1&=& c_1 \tr
  \left(\Box U^\dagger \Box U\right), \label{L1}\\
 \mathcal{L}_2&=& c_2 \tr
  \left(
   \der_\mu U^\dagger \der_\mu U \der_\nu U^\dagger \der_\nu U
  \right), \label{L2}\\
 \mathcal{L}_3&=& c_3 \tr
  \left(
   \der_\mu U^\dagger \der_\nu U \der_\mu U^\dagger \der_\nu U
  \right), \label{L3}\\
  \mathcal{L}_4&=& c_4 \tr
  \left(
   \der_\mu U^\dagger \der_\mu U 
   \right)
   \left(
   \der_\nu U^\dagger \der_\nu U
  \right), \label{L4}\\
 \mathcal{L}_5&=& c_5 \tr
  \left(
   \der_\mu U^\dagger \der_\nu U 
   \right)
   \left(
   \der_\mu U^\dagger \der_\nu U
  \right), \label{L5}\\
  \mathcal{L}_6&=& c_6 
   \big[
    \tr(\Box U^{\dagger}\partial_{\mu}U)\tr(\partial_{\mu}U^{\dagger}U)
    \nonumber \\
 &&\hspace{2em}
  {}+\tr(U^{\dagger}\partial_{\mu}U)\tr(\partial_{\mu}U^{\dagger}\Box U)
  \big]. \label{L6}
\end{eqnarray}
Expanding them in powers of $\phi$, and comparing the contributions to
the effective action with those in Eq.~(\ref{eqn:G0-2}) with the help of
the formulae given in Appendix~\ref{group}, we see that we can cancel
some of the divergences in Eq.~(\ref{eqn:G0-2}) with the following
values of $c_i$'s,
\begin{align}
 &c_{1}=0,
 &c_{2}&=\frac{N}{48(4\pi)^{2}}\bar{\varepsilon}^{-1},
 \nonumber \\
 &c_{3}=\frac{N}{96(4\pi)^{2}}\bar{\varepsilon}^{-1},
 &c_{4}&=\frac{1}{32(4\pi)^{2}}\bar{\varepsilon}^{-1},
 \nonumber \\
 &c_{5}=\frac{1}{16(4\pi)^{2}}\bar{\varepsilon}^{-1},
 &c_{6}&=\frac{1}{16(4\pi)^{2}}\bar{\varepsilon}^{-1},
\end{align}
for the case of $\overline{MS}$ renormalization scheme.
But, importantly, there are some divergences which cannot be canceled by
the symmetric counterterms with any choice of the $c_i$'s. They are
ANTs, which contribute to $B_k^{(\varphi^3)}(p)$. 

After adding the counterterms, we may define the apparently invariant
(and finite) part of the effective action $\GITs[\varphi]$ of
$\mathcal{O}(p^4)$ coming from one-loop diagrams as
\begin{widetext}
 \begin{eqnarray}
 \GITs[\varphi]
  \!&=&\!
  -\frac{1}{4F^{4}(4\pi)^{2}}
  \int\left[\prod_{i=1}^4\mom{p_{i}}\right]
  (2\pi)^{4}\delta^{4}\left(\sum_{i=1}^{4}p_{i}\right)
  \prod_{i=1}^4 \varphi_{l_{i}}(p_{i})
  \nonumber\\
  &&
  \times
  \left\{
    \left[
     2-\ln\frac{(p_{1}+p_{2})^{2}}{\mu^{2}}\right]
    (d_{kij}d_{kl_{1}l_{2}}-d_{kil_{1}}d_{kjl_{2}})
    (d_{gji}d_{gl_{3}l_{4}}-d_{gjl_{3}}d_{gil_{4}})
    (p_{1}\cdot p_{2})(p_{3}\cdot p_{4})
  \frac{\mathstrut{}}{}
  \right.
  \nonumber\\
  &&\hspace{1em}
  \left.
    -\frac{1}{3}
    \left[
     \frac{8}{3}-\ln\frac{(p_{1}+p_{2})^{2}}{\mu^{2}}\right]
    d_{kil_{1}}d_{kjl_{2}}d_{gjl_{3}}d_{gil_{4}}  
    \left[
      (p_{1}\cdot p_{4})(p_{2}\cdot p_{3})
      -
      (p_{1}\cdot p_{3})(p_{2}\cdot p_{4})
    \right]
  \right\}
  +\mathcal{O}(\varphi^{6})\; .
 \end{eqnarray}
The rest gives the apparently noninvariant (and infinite) part,
 \begin{eqnarray}
  \GANTs[\varphi]
  &=&
  -\frac{1}{16F^{4}(4\pi)^{2}}
  \int\left[\prod_{i=1}^4\mom{p_{i}}\right]
  (2\pi)^{4}\delta^{4}\left(\sum_{i=1}^{4}p_{i}\right)
  \prod_{i=1}^4 \varphi_{l_{i}}(p_{i})
  \left[\bar{\varepsilon}^{-1}+2-\ln\frac{(p_{1}+p_{2})^{2}}{\mu^{2}}\right]
  \nonumber\\
  &&
  \times
  \big[
    d_{kij}d_{kl_{1}l_{2}}d_{gji}d_{gl_{3}l_{4}}
    (p_{1}^{2}+p_{2}^{2})
    (p_{3}^{2}+p_{4}^{2})
    +
  2d_{kij}d_{kl_{1}l_{2}}
  (d_{gji}d_{gl_{3}l_{4}}-d_{gjl_{3}}d_{gil_{4}})
  (p_{1}^{2}+p_{2}^{2})
  (p_{3}\cdot p_{4})
  \nonumber\\
  &&\hspace{0em}
  +
    2d_{gji}d_{gl_{3}l_{4}}
    (d_{kij}d_{kl_{1}l_{2}}-d_{kil_{1}}d_{kjl_{2}})
    (p_{3}^{2}+p_{4}^{2})
    (p_{1}\cdot p_{2})
  \big]
  +\mathcal{O}(\varphi^{6})\; .
  \label{eqn:GK}
\end{eqnarray}
\end{widetext}
The $\mathcal{O}(\varphi^4)$ terms contribute to $B_k^{(\varphi^3)}$.

Similarly, we can calculate the $\mathcal{O}(\varphi^2 K^1)$ terms of
the effective action $\Gamma^{(1)}[\varphi, K]$. At the first order in
K, we have
\begin{eqnarray}
  &&\frac{1}{2}\ln\Det[D_{F}^{-1}]|_{K^{1}}
   \nonumber \\
  &=&\!\!
  -\frac{1}{2}\!\intmomd{k_{1}}\!\!\intmomd{k_{2}}
  \frac{1}{k_{1}^{2}}G_{0,ij}(k_{1},\!-k_{2})
  \frac{1}{k_{2}^{2}}G_{1,ji}(k_{2},\!-k_{1})
  \nonumber \\
  &&+\mathcal{O}(\varphi^4 K)\; .
   \label{first_order_in_K}
\end{eqnarray}

We obtain for the second term
\begin{eqnarray}
  &&\frac{1}{4F^{3}(4\pi)^{2}}
  \intmom{p_{1}}\intmom{p_{2}}\intmom{p}
  \nonumber \\
 &&\times 
  (2\pi)^{4}\delta^{4}(p_{1}+p_{2}+p)
  \varphi_{l_{1}}(p_{1})\varphi_{l_{2}}(p_{2})K_{k}(p)
  \nonumber\\
  &&
   \times\left(\bar{\varepsilon}^{-1}+2-\ln (p^{2}/\mu^2)\right)
   \nonumber \\
 &&
  \times
  \left[d_{gij}d_{gl_{1}l_{2}}p^{2}-2d_{gil_{1}}d_{gjl_{2}}
    (p_{1}\cdot p_{2})\right]d_{kij}\; ,
  \label{eqn:G1-2}
\end{eqnarray}
which contributes to $C_j^{(\varphi^2)}$.

Now we are ready to calculate Eq.~(\ref{cubic_in_varphi}). Noting that
we only need $B_k^{(\varphi^3)}(p)$ at $p=0$ (because of the delta
function in Eq.~(\ref{sigma0})) and that only the terms with $j=0$
survive (because of the $\delta_{0j}$) and $d_{i0k}$ is proportional to
$\delta_{ik}$ simplifies the calculation. We finally obtain for the
first term of Eq.~(\ref{cubic_in_varphi})
\begin{eqnarray}
 &&d_{ijk}\intmomd{p}
  \sigma_{j}^{(\varphi^{0})}(p)
  B_{k}^{(\varphi^{3})}(p)
  \nonumber \\
 &=&-\frac{1}{4F^{3}(4\pi)^{2}}
  \intmom{p_{1}}\intmom{p_{2}}\intmom{p_{3}}
  \nonumber \\
 &&\times
  (2\pi)^{4}\delta^{4}(p_{1}+p_{2}+p_{3})
  \varphi_{l_{1}}(p_{1})\varphi_{l_{2}}(p_{2})\varphi_{l_{3}}(p_{3})
  \nonumber\\
  &&
  \times
  \left(\bar{\varepsilon}^{-1}+2-\ln (p_{3}^{2}/\mu^2)\right) 
  d_{hnm}d_{hl_{3}i}
  \nonumber \\
 &&\times
    \left[
    d_{gmn}d_{gl_{1}l_{2}}p_{3}^{2}
    -
    d_{gml_{1}}d_{gnl_{2}}
    (p_{1}\cdot p_{2})
  \right]
  p_{3}^{2}\; .
  \label{eqn:1st-term}
\end{eqnarray}
The calculation of the second term of Eq.~(\ref{cubic_in_varphi}) is
simpler and one easily finds that it is the same as
Eq.~(\ref{eqn:1st-term}) but with the opposite sign. Thus we have
explicitly demonstrated that Eq.~(\ref{eqn:Zinn-Justin-1loop-K0-mon2})
is satisfied in $\mathcal{O}(\varphi^3)$.

We have shown that the one-loop effective action $\Gamma^{(1)}[\varphi]$
is not invariant under the naive axial-vector transformation. But it
does satisfy the WT identity so that the symmetry is preserved.

\section{Divergent part of $\GANTs[\varphi]$ to all order in $\varphi$}

In the previous section, by expanding $\Gamma[\varphi,K]$ in powers of
$\varphi$ up to including $\mathcal{O}(\varphi^4)$ and
$\mathcal{O}(\varphi^2 K^1)$, we obtain $\GANTs[\varphi]$ to
$\mathcal{O}(\varphi^4)$. But the invariance expressed by the
Zinn-Justin equation is actually more powerful; it determines the form of
the divergent part (i.e., the part that is proportional to
$\bar{\varepsilon}^{-1}$) of $\GANTs[\varphi]$ to all order of
$\varphi$. The following construction originally appears in
Ref.~\cite{Brezin:1976ap}. It is similar to the analysis done by Ferrari
and Quadri~\cite{Ferrari:2005va} for the case of $SU(2) \times SU(2)$.

Let us return to Eq.~(\ref{one-loop-ZJ}). It can be written as
\begin{equation}
  s_{i}\Gamma^{(1)}[\varphi,K]=0\; ,
\end{equation}
by introducing a differential operator $s_{i}$,
\begin{equation}
  s_{i}
  \equiv
  d_{ijk}
  \int d^{d}x
  \left(
    \sigma_{j}(x)
    \frac{\delta}{\delta \varphi_{k}(x)}
    -H_{k}(x)
    \frac{\delta}{\delta K_{j}(x)}
  \right)\; ,
\end{equation}
where $H_{k}$ is defined as
\begin{equation}
  H_{k}(x)\equiv -\frac{\delta S[\varphi,K]}{\delta \varphi_{k}(x)}
  =-\frac{\delta S[\varphi]}{\delta\varphi_{k}(x)}
  -\int\! d^d y
  \frac{\partial\sigma_{i}(y)}{\partial\varphi_{k}(x)}K_{i}(y)
  \; .
\end{equation}
 Since
$\GITs[\varphi]$ satisfies
\begin{equation}
 s_{i}\GITs[\varphi]=0\; ,
\end{equation}
we have 
\begin{equation}
 s_{i}\left(\GANTs[\varphi]+\GK[\varphi,K]\right)=0\; .
\end{equation}

In the following, we discuss only the divergent parts, that is, the
\textit{local} functionals of $\varphi$ and $K$.

An important observation is that $K$ and $H$ transform in the same way
as $\sigma$ and $\varphi$ do :
\begin{equation}
  \left\{
    \begin{array}{l}
      s_{i}K_{j}=-d_{ijk}H_{k}\; ,\\
      s_{i}H_{j}=d_{ijk}K_{k}\; ,
    \end{array}
  \right.    
\end{equation}
\begin{equation}
  \left\{
    \begin{array}{l}
      s_{i}\sigma_{j}=-d_{ijk}\varphi_{k}\; ,\\
      s_{i}\varphi_{j}=d_{ijk}\sigma_{k}\; .
    \end{array}
  \right.
\end{equation}
Invariants are most easily formed in terms of $U=(\sigma
+i\varphi)/F$. In the same way, we introduce
\begin{equation}
 L=\frac{1}{F}\left(K+iH\right)
  =\frac{2}{F}\left(K_{i}+iH_{i}\right)T_{i}\; ,
  \label{defL}
\end{equation}
and form invariants in terms of $U$ and $L$. Note that $L$ has the same
property under parity as that of $U$.

We remark that the divergence of the effective action occurs in the
terms at most quadratic in $K$ in the one-loop approximation. To
represent them, therefore, we only consider the invariants at most
quadratic in $L$. We also remark that the vertices containing $K$ do
not involve derivatives, thus the divergent part of the effective action
quadratic in $L$ does not involves derivatives, either.

There are five independent invariants contributing to $\GANTs[\varphi] +
\GK[\varphi,K]$ up to including $\mathcal{O}(p^4)$ and
$\mathcal{O}(\varphi^4)$ when expanded and consisting of $U$ and $L$,
\begin{eqnarray}
  O_{1}
  &=&
  \tr(L^{\dag}L) , \label{O1}
  \\ 
  O_{2}
  &=&
  \frac{1}{4}\left[
    \tr(U^{\dag}L)+\tr(L^{\dag}U)
  \right]^{2} , \label{O2}
  \\
  O_{3}
  &=&
  \frac{1}{2}\left[
    \tr(\Box U^{\dag}L)+\tr(L^{\dag}\Box U)
  \right] , \label{O3}
  \\
  O_{4}
  &=&
  \frac{1}{2}\tr(\partial_{\mu}U^{\dag}\partial_{\mu}U)
  \left[
    \tr(U^{\dag}L)+\tr(L^{\dag}U)
  \right] , \label{O4}
  \\
  O_{5}
  &=&
  \frac{1}{4}
  \left[\tr(U^{\dag}\partial_{\mu}U)-\tr(\partial_{\mu}U^{\dag}U)\right]
  \nonumber \\
 &&\hspace{1em}\times{}
  \left[\tr(\partial_{\mu}U^{\dag}L)-\tr(L^{\dag}\partial_{\mu}U)\right],
  \label{O5}
\end{eqnarray}
besides the ones consisting only of $U$ listed in Eqs.~(\ref{L1}) --
(\ref{L6}).

One might think that there should be more invariants which start with
$\varphi^6$ when expanded, but because of the constraint $U^\dagger
U=I$, such an invariant requires at least six derivatives; there is no
other invariant than $O_i\ (i=1,\cdots, 5)$ to this order. Thus
$\GANTs[\varphi] + \GK[\varphi,K]$ can be expressed as a linear
combination of the integrals of these invariants.

 By comparing the $\mathcal{O}(\varphi^4)$ and $\mathcal{O}(\varphi^2
K)$ terms with Eqs.~(\ref{eqn:GK}) and (\ref{first_order_in_K}), we can
determine the coefficients. We can determine the divergent part of
$\GANTs[\varphi] + \GK[\varphi,K]$ to all order in $\varphi$ as,
\begin{eqnarray}
 &&\left(\GANTs[\varphi] + \GK[\varphi,K]\right)_{\mbox{\scriptsize div}}
  \nonumber \\
  &=&
  -\frac{\bar{\varepsilon}^{-1}}{8(4\pi)^{2}}
  \int\! d^{d}x
  \left[
    N(O_{1}-O_{3})
    +O_{2}+O_{4}-2O_{5}
  \right].
  \nonumber \\
  \label{eqn:ANTs}
\end{eqnarray}

By setting $K=0$, we see that $L$ is then proportional to $\delta
S/\delta\varphi$,
\begin{equation}
 L\rightarrow -\frac{2i}{F}\frac{\delta S}{\delta\varphi_i}T_{i}\; .
\end{equation}
We therefore see that the divergent part of $\GANTs[\varphi]$ is
proportional to $\delta S/\delta\varphi$.

It is well known~\cite{'tHooft:1973us} that the part of the one-loop
effective action that is proportional to $\delta S/\delta\varphi$ does not
contribute to the S-matrix elements, due to the ``equivalence
theorem~\cite{Kamefuchi:1961sb}'' That is, ANTs of the amputated
connected Green function (ACGF) vanish on shell.

\section{Six-point functions}
\label{six-point}

There seems to be a little confusion about the ``use of the equations of
motion,'' $\delta S/\delta \varphi=0$, and the ``on-shell condition,''
$p_i^2=0$. Some authors use them interchangeably. But they are
essentially different.  ANTs of the one-loop effective action vanish
by the use of the equations of motion. 

One might think that it implies that the ANTs contained in the $n$-point
proper vertices generated by the effective action also vanish \textit{on
shell}. But this is not the case.

It is not the $n$-point proper vertices, but the ACGF whose ANTs vanish
on shell, because of the equivalence theorem.

For the four-point functions, however, the difference is not seen; the
proper vertex coincides with the ACGF in the one-loop
approximation. Thus ANTs of the proper vertex happen to vanish on
shell. See Eq.~(\ref{eqn:GK}).

It is therefore interesting to examine the six-point functions because
the difference between the ACGF and the proper vertex first appears in
the six-point functions at the one-loop order. Since we have obtained
the divergent part of the one-loop effective action to all order in
$\varphi$, we are ready to examine the $n$-point functions with $n\ge
6$.

\begin{figure}[t]
 \includegraphics[width=0.8\linewidth,clip]{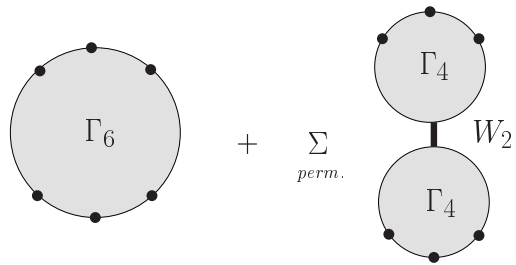}
\caption{\label{sixpointdots}The graphical representation of the right
hand side of Eq.~(\ref{Ws_and_Gammas}). A solid circle stands an
amputated external line and the solid line for the full propagator,
 $W_{\rho\sigma}$, abbreviated as $W_2$. The
gray blobs stand for the proper vertices. $\Gamma_6$ is the abbreviated
notation for $\Gamma_{\alpha_1\cdots\alpha_6}$, and $\Gamma_{4}$ for
$\Gamma_{\alpha_{i_1}\alpha_{i_2}\alpha_{i_3}\rho}$ and
$\Gamma_{\alpha_{i_4}\alpha_{i_5}\alpha_{i_6}\rho}$. The sum is taken
 over the permutation of the external lines.}
\end{figure}

We first show that the ANTs of the six-point proper vertex do not vanish
on shell.  Because we are interested in the proper vertices in the
absence of $K$, we set $K=0$. By expanding $O_i\ (i=1,\cdots,
5)$ in terms of $\varphi$, we see that the contributions from
$O_2$, $O_3$ vanish on shell. The
$O(\varphi^6)$ part of $O_1$ is given by
\begin{equation}
  \frac{1}{F^{4}}
  d_{gji_{1}}d_{gi_{2}i_{3}}d_{hji_{4}}d_{hi_{5}i_{6}}
  \varphi_{i_{1}}\varphi_{i_{4}}
  \partial_{\mu}\varphi_{i_{2}}\partial_{\mu}\varphi_{i_{3}}
  \partial_{\nu}\varphi_{i_{5}}\partial_{\nu}\varphi_{i_{6}},
\end{equation}
up to the terms proportional to $\Box\varphi_i$, which vanish on
shell. Similarly, from $O_4$ and $O_5$, we have the
contributions that do not vanish on shell,
\begin{eqnarray}
 &&\frac{1}{F^{2}}
  d_{ijk}d_{ilm}
  \varphi_{j}\varphi_{k}\partial_{\mu}\varphi_{l}\partial_{\mu}\varphi_{m}
  (\partial_{\nu}\varphi_{g})^{2}, 
  \\
 &&
  \frac{1}{F^{2}}\sqrt{\frac{N}{2}}
  d_{i,jk,lm}
  \varphi_{j}\partial_{\mu}\varphi_{k}
  \varphi_{i}
  \partial_{\nu}\varphi_{l}\partial_{\nu}\varphi_{m}
  \partial_{\mu}\varphi_{0},
  \nonumber \\
\end{eqnarray}
respectively, where $d_{i,jk,lm}$ stands for $d_{igh} d_{gjk}
d_{hlm}$. Substituting these contributions in Eq.~(\ref{eqn:ANTs}), we
see that the divergent part of $\GANTs[\varphi]$ contains
\begin{eqnarray}
 &&-\frac{\bar{\varepsilon}^{-1}}{8F^{6}(4\pi)^{2}}
  \int \left[\prod_{i=1}^{6}\frac{d^{4}p_i}{(2\pi)^{4}}\right]
  \!\!(2\pi)^{4}\delta^{4}\!\left(\sum_{i=1}^{6}p_i\!\right)
  \prod_{k=1}^{6} \varphi_{i_{k}}(p_{k})
  \nonumber\\
  &&
  \times
  d_{i_{1}i_{2},i_{3}i_{4},i_{5}i_{6}}(p_{1}\cdot p_{2})(p_{3}\cdot
  p_{4})
  \label{ants6}
\end{eqnarray}
that gives rise to the nonvanishing contribution to the six-point proper
vertex. Here we have introduced another notation,
\begin{eqnarray}
 &&d_{i_{1}i_{2},i_{3}i_{4},i_{5}i_{6}}
  \nonumber \\
  &=&
  Nd_{gi_{1}i_{2}}d_{hi_{3}i_{4}}
  (d_{gki_{5}}d_{hki_{6}}+d_{gki_{5}}d_{hki_{6}})
  \nonumber \\
 &&{}+2d_{ki_{5}i_{6}}
  (d_{ki_{1}i_{2}}\delta_{i_{3}i_{4}}+d_{ki_{3}i_{4}}\delta_{i_{1}i_{2}})
  \nonumber\\
  &&
  -\sqrt{\frac{N}{2}}
  \big[
  (d_{i_{5},i_{1}i_{2},i_{4}i_{6}}+d_{i_{6},i_{1}i_{2},i_{4}i_{5}})
  \delta_{i_{3}0}
  \nonumber \\
 &&\hspace{3em}
  +
  (d_{i_{5},i_{1}i_{2},i_{3}i_{6}}+d_{i_{6},i_{1}i_{2},i_{3}i_{5}})
  \delta_{i_{4}0}  
  \nonumber\\
  &&\hspace{3em}
  +
  (d_{i_{5},i_{3}i_{4},i_{2}i_{6}}+d_{i_{6},i_{3}i_{4},i_{2}i_{5}})
  \delta_{i_{1}0}
  \nonumber\\
  &&\hspace{3em}
  +
  (d_{i_{5},i_{3}i_{4},i_{1}i_{6}}+d_{i_{6},i_{3}i_{4},i_{1}i_{5}})
  \delta_{i_{2}0}
  \big].
\end{eqnarray}
Thus, we have established that the divergent part of ANTs in the
six-point proper vertex does not vanish on shell.

We will show that ANTs contained in the six-point ACGF do vanish on
shell. In an abbreviated notation, the six-point ACGF is written as
\begin{eqnarray}
  &&
  \Gamma_{\alpha_{1}\beta_{1}}
  \Gamma_{\alpha_{2}\beta_{2}}
  \Gamma_{\alpha_{3}\beta_{3}}
  \Gamma_{\alpha_{4}\beta_{4}}
  \Gamma_{\alpha_{5}\beta_{5}}
  \Gamma_{\alpha_{6}\beta_{6}}
  W_{\beta_{1}\beta_{2}\beta_{3}\beta_{4}\beta_{5}\beta_{6}}
  \nonumber\\
  &=&
  \Gamma_{\alpha_{1}\alpha_{2}\alpha_{3}\alpha_{4}\alpha_{5}\alpha_{6}}
  \nonumber\\
  &&
  +
  \left(
    \Gamma_{\alpha_{1}\alpha_{2}\alpha_{3}\rho}
    \Gamma_{\alpha_{4}\alpha_{5}\alpha_{6}\sigma}
    +
    \mbox{9 similar terms}
  \right)W_{\rho\sigma},
\label{Ws_and_Gammas}
\end{eqnarray}
where $W_{\alpha_{1}\alpha_{2}\cdots\alpha_{n}}$ is a connected Green
function,
\begin{equation}
 W_{\alpha_{1}\alpha_{2}\cdots\alpha_{n}}
  \equiv
  (-1)^n
  \left.
   \frac{\delta^{n}W[J]}{\delta J_{\alpha_{1}}
   \cdots \delta J_{\alpha_{n}}}
  \right|_{J=0},
\end{equation}
where $W[J]=W[J,K=0]$ and
$\Gamma_{\alpha_{1}\alpha_{2}\cdots\alpha_{n}}$ is a proper vertex,
\begin{equation}
 \Gamma_{\alpha_{1}\alpha_{2}\cdots\alpha_{n}}
  \equiv
  \left.
   \frac{\delta^{n}\Gamma[\varphi]}
   {\delta \varphi_{\alpha_{1}}\cdots\delta \varphi_{\alpha_{n}}}
  \right|_{\varphi=0}.
\end{equation}
Note that because the invariant terms are made finite by the addition of
the invariant counterterms, the divergent parts of the both sides of
Eq.~(\ref{Ws_and_Gammas}) come only from ANTs.

The assertion is that, despite the nonvanishing divergent part of ANTs
in one-loop contributions to the first term in
Eq.~(\ref{Ws_and_Gammas}) on shell, the sum does not contain such ANTs.
Note that the tree contributions of course do not contain ANTs.

In the one-loop approximation, the propagator $W_{\rho\sigma}$ is the
free one thanks to a property of dimensional regularization.  In the
product $\Gamma_{\alpha_{1}\alpha_{2}\alpha_{3}\rho}
\Gamma_{\alpha_{4}\alpha_{5}\alpha_{6}\sigma}$ one of the two $\Gamma$'s
must be the one-loop contribution while the other must be the tree
one. Thus the noninvariant divergent contributions to the
$\Gamma_{\alpha_{1}\alpha_{2}\alpha_{3}\rho}
\Gamma_{\alpha_{4}\alpha_{5}\alpha_{6}\sigma}W_{\rho\sigma}$ in the
one-loop approximation contain the following divergent part,
\begin{widetext}
\begin{eqnarray}
  &&
   \intmomd{p}\intmomd{k}
    \bigg[
    (2\pi)^{4d}
    \frac{\delta^{4}S[\varphi]}
    {\delta\varphi_{i_{1}}(q_{1})
    \delta\varphi_{i_{2}}(q_{2})
    \delta\varphi_{i_{3}}(q_{3})
    \delta\varphi_{j}(p)}
    (2\pi)^{4d}
    \frac{\delta^{4}\left(\GANTs[\varphi]\right)_{\mbox{\scriptsize div.}}}
    {\delta\varphi_{i_{4}}(q_{4})
    \delta\varphi_{i_{5}}(q_{5})
    \delta\varphi_{i_{6}}(q_{6})
    \delta\varphi_{k}(k)}
    \nonumber\\
  &&\hspace{3em}
    +   
    (2\pi)^{4d}
    \frac{\delta^{4}\left(\GANTs[\varphi]\right)_{\mbox{\scriptsize div.}}}
    {\delta\varphi_{i_{1}}(q_{1})
    \delta\varphi_{i_{2}}(q_{2})
    \delta\varphi_{i_{3}}(q_{3})
    \delta\varphi_{j}(p)}
    (2\pi)^{4d}
    \frac{\delta^{4}S[\varphi]}
    {\delta\varphi_{i_{4}}(q_{4})
    \delta\varphi_{i_{5}}(q_{5})
    \delta\varphi_{i_{6}}(q_{6})
    \delta\varphi_{k}(k)}
    \bigg]_{\varphi=0}
  \frac{-1}{k^{2}}(2\pi)^{d}\delta^{d}(k+p)\delta_{jk}
  \nonumber\\
  &=&
  \frac{2\bar{\varepsilon}^{-1}}{F^{6}(4\pi)^{2}}
  (2\pi)^{d}\delta^{d}\left( \sum_{i=1}^6q_{i}\right)
  \bigg[
    d_{i_{1}i_{2}i_{3},i_{4}i_{5}i_{6}}(q_{1}\cdot q_{2})(q_{4}\cdot q_{5})
    +d_{i_{1}i_{2}i_{3},i_{4}i_{6}i_{5}}(q_{1}\cdot q_{2})(q_{4}\cdot q_{6})
    +d_{i_{1}i_{2}i_{3},i_{5}i_{6}i_{4}}(q_{1}\cdot q_{2})(q_{5}\cdot q_{6})
  \nonumber\\
  &&\hspace{3em}
  +d_{i_{1}i_{3}i_{2},i_{4}i_{5}i_{6}}(q_{1}\cdot q_{3})(q_{4}\cdot q_{5})
  +d_{i_{1}i_{3}i_{2},i_{4}i_{6}i_{5}}(q_{1}\cdot q_{3})(q_{4}\cdot q_{6})
  +d_{i_{1}i_{3}i_{2},i_{5}i_{6}i_{4}}(q_{1}\cdot q_{3})(q_{5}\cdot q_{6})
  \nonumber\\
  &&\hspace{3em}
    +d_{i_{2}i_{3}i_{1},i_{4}i_{5}i_{6}}(q_{2}\cdot q_{3})(q_{4}\cdot q_{5})
    +d_{i_{2}i_{3}i_{1},i_{4}i_{6}i_{5}}(q_{2}\cdot q_{3})(q_{4}\cdot q_{6})
    +d_{i_{2}i_{3}i_{1},i_{5}i_{6}i_{4}}(q_{2}\cdot q_{3})(q_{5}\cdot q_{6})
  \bigg]
  \nonumber \\
 &&{}+(\mbox{terms those do not contribute on shell}),
\end{eqnarray}
\end{widetext}
where we introduce
\begin{eqnarray}
 &&d_{i_{1}i_{2}i_{3},i_{4}i_{5}i_{6}}
  \nonumber \\
 &\equiv&
  d_{gij}d_{li_{1}i_{2}}d_{i_{3}lk}
  (d_{hij}d_{hi_{4}i_{5}}-d_{hii_{4}}d_{hji_{5}})
  d_{i_{6}gk}
  \nonumber \\
 &&{}+
  (i_{1}i_{2}i_{3}\leftrightarrow i_{4}i_{5}i_{6}).
\end{eqnarray}
Note that, with the help of the formulae given in
Appendix~\ref{sec:su2}, one can show that 
\begin{eqnarray}
 d_{i_{1}i_{2},i_{3}i_{4},i_{5}i_{6}}
  &=&
  d_{i_{1}i_{2}i_{5},i_{3}i_{4}i_{6}}
  +
  d_{i_{1}i_{2}i_{6},i_{3}i_{4}i_{5}}\; ,
  \label{d_and_d}
\end{eqnarray}
which is useful in the following demonstration.

In a similar way, we can calculate all the divergent terms contained in 
\begin{equation}
 \left(
    \Gamma_{\alpha_{1}\alpha_{2}\alpha_{3}\rho}
    \Gamma_{\alpha_{4}\alpha_{5}\alpha_{6}\sigma}
    +
    \mbox{9 similar terms}
 \right)W_{\rho\sigma}
\end{equation}
at one-loop that do not vanish on shell. There are 90 similar terms to
\begin{equation}
 \frac{2\bar{\varepsilon}^{-1}}{F^{6}(4\pi)^{2}}
  d_{i_{1}i_{2}i_{3},i_{4}i_{5}i_{6}}(q_{1}\cdot q_{2})(q_{4}\cdot q_{5}).
\end{equation}
On the other hand, the differentiation of Eq.~(\ref{ants6}) with respect
to $\varphi$ six times produces $6!=720$ terms but every 16 ($=2^4$)
terms are actually identical. (Note the symmetry under the interchange
of the indices, $i_1\leftrightarrow i_2$, $i_3\leftrightarrow i_4$,
$i_5\leftrightarrow i_6$, and $(i_1i_2) \leftrightarrow (i_3i_4)$ of the
representative term $d_{i_{1}i_{2},i_{3}i_{4},i_{5}i_{6}}(p_{1}\cdot
p_{2})(p_{3}\cdot p_{4})$.) Thus we have 45 ($=720/16$) similar terms to
\begin{equation}
 -\frac{2\bar{\varepsilon}^{-1}}{F^{6}(4\pi)^{2}}
  d_{i_{1}i_{2},i_{3}i_{4},i_{5}i_{6}}
  (q_{1}\cdot q_{2})(q_{3}\cdot q_{4}).
\end{equation}
Now substituting Eq.~(\ref{d_and_d}), we see that the divergent terms in
the first term of the right hand side of Eq.~(\ref{Ws_and_Gammas}) is
exactly equal to those in the second term with the opposite sign. Thus
we have shown that the divergent terms cancel in the ACGF on shell.

\section{Summary}
\label{summary}

In this paper, we have explained why ANTs emerge in theories with
nonlinearly realized symmetry.  The naive variation of the classical
field $\varphi$ under the symmetry is in general not equal to the
expectation value of the variation of the quantum field $\phi$, so that
the effective action $\Gamma[\varphi]$ is not invariant under the naive
variation of $\varphi$. In this respect, the emergence of ANTs is not
limited to NLSMs. 

We have formulate the WT identity as an inhomogeneous Zinn-Justin
equation for the $U(N) \times U(N)$ NLSMs to investigated the nature of
ANTs. We have shown that ANTs are consistent with the symmetry.

An explicit one-loop calculation is presented for the four-point
functions. We then determine the divergent part of the ANTs in the
one-loop effective action, $\GANTs[\varphi]$, to all order in $\varphi$.

It is shown that the divergent part of the ANTs in the six-point proper
vertex does not vanish when the external momenta are all set to be on
shell, in contrast to the four-point proper vertex, whose divergent part
of the ANTs vanishes on shell. We have shown that, despite its presence
in the proper vertices, ANTs contained in the six-point ACGF do vanish
on shell. We believe that higher-point ACGF also have this property.

Note that we do not subtract divergences in $\GANTs[\varphi]$; the
introduction of noninvariant counterterms cause the violation of the
original symmetry. Even though they are divergent, ANTs do not
contribute to the S-matrix elements after all. It provides an example
where the effective action is divergent while the S-matrix elements are
finite.

We expect that similar properties hold even in higher order in the
loop-wise expansion.

\appendix
\section{Some useful formulae for $U(N)$}
\label{group}

In this Appendix, we summarize some formulae for the symmetric tensor
$d_{ijk}$, which are used in the loop calculations. See
Refs.~\cite{MacFarlane:1968vc, Geddes:1978ji} for the details.

We consider the $U(N)$ generators as $N\times N$ matrices, $T_i\
(i=0,1,\cdots, N^2-1)$, satisfying the relations in
Eqs.~(\ref{U_N_algebra}), (\ref{T0}), and (\ref{f_and_d}).

By considering the Jacobi identities,
\begin{eqnarray}
 &&{}[[T_i,T_j],T_k]+[[T_j,T_k],T_i]+[[T_k,T_i],T_j]=0,\\
 &&{}[\{T_i, T_j\},T_k]+[\{T_j, T_k\},T_i]+[\{T_k, T_i\},T_j]=0, \qquad
  \\
 &&{}[T_k,[T_i,T_j]]=\{T_j,\{T_k,T_i\}\}-\{T_i,\{T_j,T_k\}\},
\end{eqnarray}
we find the following relations between $f_{ijk}$'s and $d_{ijk}$'s,
\begin{eqnarray}
    &&f_{ilm}f_{mjk}+f_{jlm}f_{imk}+f_{klm}f_{ijm}=0,\\
    &&f_{ilm}d_{mjk}+f_{jlm}d_{imk}+f_{klm}d_{ijm}=0,\\
    &&f_{ijm}f_{klm}-d_{ikm}d_{jlm}+d_{jkm}d_{ilm}=0.
\end{eqnarray}
From them one can derive useful relations.

Let us introduce $N^2 \times N^2$ matrices, $(D_{j})_{ik}$,
\begin{equation}
    (D_{j})_{ik}\equiv d_{ijk},
\end{equation}
which satisfy
\begin{eqnarray}
 \tr(D_{i}D_{j})&=&N(\delta_{ij}+\delta_{i0}\delta_{j0}),
  \\
 \tr(D_{i}D_{j}D_{k})
    &=&
    \frac{N}{2}d_{ijk}
    +\sqrt{\frac{N}{2}}
    (
     \delta_{i0}\delta_{jk}
     +\delta_{j0}\delta_{ik}
     \nonumber \\
 &&\hspace{6em}+\delta_{k0}\delta_{ij}),
  \\
 \tr(D_{i}D_{j}D_{k}D_{l})
    &=&
    \frac{N}{4}(d_{ijm}d_{mkl}-f_{ijm}f_{mkl})
    \nonumber \\
 &&{}+\frac{1}{2}
  (\delta_{ij}\delta_{kl}+\delta_{ik}\delta_{jl}+\delta_{il}\delta_{jk})
    \nonumber\\
    &&
    +\frac{\sqrt{2N}}{4}
    (\delta_{i0}d_{jkl}+\delta_{j0}d_{ikl}+\delta_{k0}d_{ijl}
    \nonumber \\
 &&{}\hspace{4em}+\delta_{l0}d_{ijk}).
\end{eqnarray}
They are used in reducing the result of Eq.~(\ref{eqn:G0-2}) in a
compact form for the comparison with the local terms listed in
Eqs.(\ref{L1}) through (\ref{L6}).

\section{$SU(2) \times SU(2)$ case}
\label{sec:su2}

In this Appendix, we present the results for the $SU(2) \times SU(2)$
case. Because of the extra $U(1)$ factors, the $SU(2) \times SU(2)$
results cannot be obtained directly from those for $U(N) \times
U(N)$. The purpose of this Appendix is to provide a useful reference for
those who are familiar with the literature devoted to the $SU(2) \times
SU(2)$ case.

For the $SU(2) \times SU(2)$ case, the most useful parameterization is
given by Eq.~(\ref{su2sqrt}). (But we continue to use the notation
$\phi^a$ instead of $\pi^a$.) We consider the generating functional,
\begin{equation}
 e^{-W[J, K]}\!=\!\int\! \mathcal{D}\phi \;
  \exp
  \left[
   -\!\int\! d^4x \left(\mathcal{L} +\sigma K\right)
   \!+\!\int\! d^4x \phi_{a}J_{a}
  \right],
\end{equation}
and define the effective action as usual. The inhomogeneous Zinn-Justin
equation is given by
\begin{equation}
 \int d^4x 
  \left(
   \frac{\delta \Gamma[\varphi, K]}{\delta K(x)}
   \frac{\delta \Gamma[\varphi, K]}{\delta \varphi_a(x)}
   +\varphi_a(x) K(x)
  \right)=0.
\end{equation}

The one-loop calculation for the effective action,
$\Gamma^{(1)}[\varphi, K]$, can be done in a similar way, but the
results are much simpler.  The $K$-independent part contains
\begin{eqnarray}
 &&-\frac{\bar{\varepsilon}^{-1}}{16F^{4}(4\pi)^{2}}
  \!\int \!\left[\prod_{i=1}^4\frac{d^{4} p_i}{(2\pi)^{4}}\right]
  \!(2\pi)^{4}\delta^{4}\!\left(\sum_{i=1}^4p_{i}\!\right)
  \nonumber\\
 &&{} \times \varphi_{a}(p_{1})\varphi_{a}(p_{2})
  \varphi_{b}(p_{3})\varphi_{b}(p_{4})
  \nonumber \\
 &&{}
  \times
  \bigg[
    3(p_{1}^{2}+p_{2}^{2})
    (p_{3}^{2}+p_{4}^{2})
    +
    4(p_{1}^{2}+p_{2}^{2})
    (p_{3}\cdot p_{4})
    \nonumber \\
 &&\hspace{1em}{}
    +
    4(p_{3}^{2}+p_{4}^{2})
    (p_{1}\cdot p_{2})
    +
    \frac{8}{3}(p_{1}\cdot p_{2})
    (p_{3}\cdot p_{4})
    \nonumber \\
 &&\hspace{1em}{}
    +
    \frac{8}{3}(p_{1}\cdot p_{4})
    (p_{2}\cdot p_{3})
    +
    \frac{8}{3}(p_{1}\cdot p_{3})
    (p_{2}\cdot p_{4})
  \bigg]
  \nonumber\\
  &&\hspace{1em}{}
  +\text{(finite terms)},
  \label{su2K0}
\end{eqnarray}
instead of Eq.~(\ref{eqn:G0-2}). There are only three independent
symmetric counterterms of this order because of the relations,
\begin{eqnarray}
 \tr(\partial_{\mu}U^{\dag}\partial_{\mu}U\partial_{\nu}U^{\dag}\partial_{\nu}U)
  &=&
  \frac{1}{2}[\tr(\partial_{\mu}U^{\dag}\partial_{\mu}U)]^{2},
  \label{eqn:SU(2)-trace1}
  \\
 \tr(
  \partial_{\mu}U^{\dag}\partial_{\nu}U\partial_{\mu}U^{\dag}\partial_{\nu}U
  )
  &=&
  [\tr(\partial_{\mu}U^{\dag}\partial_{\nu}U)]^{2}
  \nonumber \\
 &&{}
  -\frac{1}{2}[\tr(\partial_{\mu}U^{\dag}\partial_{\mu}U)]^{2},\ \ 
  \label{eqn:SU(2)-trace2}
\end{eqnarray}
which hold for the $SU(2)$ case, and
\begin{equation}
 \tr(U^\dagger \der_\mu U)=\tr(\der_\mu U^\dagger U)=0, \label{su2trace}
\end{equation}
due to the tracelessness of the generators. Thus there are only
\textit{three} independent invariants, $\mathcal{L}_1$, $\mathcal{L}_4$,
and $\mathcal{L}_5$. By expanding $\mathcal{L}_{i}\ (i=1,4,5)$ in
$\varphi$ and comparing the momentum dependence of them with the terms
in Eq.~(\ref{su2K0}), we see that some of the divergences can be removed
by these counterterms with the coefficients,
\begin{equation}
 c_1=0, \quad
  c_{4}=\frac{1}{24(4\pi)^{2}}\bar{\varepsilon}^{-1}, \quad
  c_{5}=\frac{1}{12(4\pi)^{2}}\bar{\varepsilon}^{-1},
\end{equation}
which agree with the result Eq.~(3.4) of
Ref.~\cite{Appelquist:1980ae}. (Note that there is a factor 2 difference
coming from the difference between their $\epsilon =4-d$ and our
$\varepsilon =(4-d)/2$, and the difference of signs coming from that we
are working in Euclidean space.)

Similarly, for the linear part in $K$, we have
\begin{eqnarray}
 &&
  \frac{1}{4F^{3}(4\pi)^{2}}
  \intmom{p_{1}}\intmom{p_{2}}\intmom{p} 
  \nonumber \\
 &&\times (2\pi)^{4}\delta^{4}(p_{1}+p_{2}+p)
  \varphi_{a}(p_{1})\varphi_{a}(p_{2})K(p)
  \nonumber\\
  &&
  \times(\bar{\varepsilon}^{-1}+2-\ln (p^{2}/\mu^2))
  \left[3p^{2}-2(p_{1}\cdot p_{2})\right],
\end{eqnarray}
instead of Eq.~(\ref{eqn:G1-2}). The divergent part of $\GANTs[\varphi]
+ \GK[\varphi,K]$ up to including $\mathcal{O}(p^4)$ may be written as a
linear combination of $O_{i}$'s listed in Eqs.~(\ref{O1}) -- (\ref{O5}),
where $L$ is now defined as
\begin{equation}
 L=\frac{1}{F}\left[K + i \vec{\tau}\cdot\vec{H}\right],
\end{equation}
where
\begin{equation}
 H_{a}\equiv -\frac{\delta S[\varphi]}{\delta \varphi_{a}}
  +\frac{\varphi_{a}}{\sigma}K,
  \label{su2H}
\end{equation}
instead of Eq.~(\ref{defL}). 

Because of the difference between $U(N)$ and $SU(2)$, however, not all
of $O_{i}$'s are independent in the case of $SU(2)$. First of all, $O_5$
vanishes because of Eq.~(\ref{su2trace}). There are also relations among
the rest.

Consider $O_1$ and $O_2$ for example. Using
(\ref{su2H}), we can rewrite them as
\begin{eqnarray}
 O_1&=&
  \frac{2}{F^2}
  \left[
   \frac{F^{2}K^{2}}{\sigma^{2}}
   -2\frac{K}{\sigma}\varphi_{a}S_{a}
   +S_{a}^{2}
  \right],
  \\
  O_2&=&
   \frac{4}{F^4}
   \left[
    \frac{F^{4}K^{2}}{\sigma^{2}}
    -2\frac{F^{2}K}{\sigma}\varphi_{a}S_{a}
    +\left(\varphi_{a}S_{a}\right)^{2}
   \right],\ \ 
\end{eqnarray}
where we have introduced the notation $S_{a}\equiv \delta
S[\varphi]/\delta \varphi_{a}$. The combination
\begin{equation}
 O_1-\half O_2=
  \frac{2}{F^4}\left[F^2S_{a}^2-(\varphi_{a}S_{a})^2\right]
\end{equation}
is independent of $K$. In fact, they are invariant under the
axial-vector transformation $SU(2)_A$ as we are going to show. Note that
the infinitesimal generator now takes the form
\begin{equation}
 s_{a}\equiv \int d^{d}x
  \left(
   \sigma(x)\frac{\delta}{\delta \varphi_a(x)}
   -H_{a}(x)\frac{\delta}{\delta K(x)}
  \right).
\end{equation}
Applications of $s_{a}$ to $S_{b}$ and $(\varphi_{b}S_{b})$ give
\begin{equation}
 s_{a}S_{b}=\frac{\varphi_{b}}{\sigma}S_{a},
  \quad 
  s_{a}(\varphi_{b}S_{b})=\frac{F^2}{\sigma}S_{a},
\end{equation}
so that
\begin{equation}
 s_{a}\left(
       O_1-\half O_2
      \right)=0.
\end{equation}
It means that, modulo invariant terms consisting only of $U$, $O_1$ is
equivalent to $O_2/2$.

Similarly we can rewrite $O_3$ and $O_4$ as
\begin{eqnarray}
 O_3&=&\frac{2}{F^2}
  \left[
   \left(\sigma\Box\sigma+\varphi_{a}\Box\varphi_{a}\right)\frac{K}{\sigma}
   -S_{a}\Box\varphi_{a}
  \right], \\
 O_4&=&\frac{2}{F^4}
  \left[(\der_\mu \sigma)^2+(\der_\mu\varphi_{a})^2\right]\!
  \left[\frac{F^2K}{\sigma}-\varphi_{a}S_{a}\right], \ \ \ \ 
\end{eqnarray}
thus the combination 
\begin{equation}
 O_3+O_4=\frac{2}{F^4}
  \left[
   -F^2S_{a}\Box\varphi_{a}
   +
   \left(
    (\der_\mu \sigma)^2+(\der_\mu\varphi_{a})^2
   \right)\varphi_{b}S_{b}
  \right]
\end{equation}
is $K$ independent and is invariant under $SU(2)_A$,
\begin{equation}
 s_{a}\left(O_3+O_4\right)=0.
\end{equation}
It means that, modulo invariant terms consisting only of $U$, $O_3$ is
equivalent to $-O_4$. Thus we have shown that there are only
\textit{two} independent ones, $O_2$ and $O_4$, for example.

The divergent terms of $\GANTs[\varphi] + \GK[\varphi,K]$ can be
expressed as a linear combination of the independent ones,
\begin{eqnarray}
 &&\left(
    \GANTs[\varphi] + \GK[\varphi,K]
   \right)_{\mbox{\scriptsize div.}}
  \nonumber \\
 &=&-\frac{1}{16(4\pi)^2}\bar{\varepsilon}^{-1}\int d^{d}x
  \left[
   3O_{2}+8O_{4}
  \right],
\end{eqnarray}
which corresponds to Eq.~(\ref{eqn:ANTs}). This agrees with the results
in the literature~\cite{Tataru:1975ys, Appelquist:1980ae, Ferrari:2005va}.

\begin{acknowledgments}
 The authors are grateful to N.~Hattori for the discussions in the early
 stage of the present investigation.
\end{acknowledgments}

\bibliography{NLS,EFT,NEFT}

\end{document}